\documentclass[a4paper]{article}

\usepackage{INTERSPEECH2020}

\usepackage{float}
\usepackage{multirow}
\usepackage{diagbox}

\title{Voice Activity Detection Scheme by Combining DNN Model with GMM Model}
\name{Lu Ma, Xiaomeng Zhang, Pei Zhao, Tengrong Su}
\address{Haier Smart Home Co., Ltd.}
\email{malu@haier.com, iamroad@163.com}

\begin{document}

\maketitle
\begin{abstract}
  Due to the superior modeling ability of deep neural network (DNN), it is widely used in voice activity detection (VAD). However, the performance may degrade if no sufficient data especially for practical data could be used for training, thus, leading to inferior ability of adaption to environment. Moreover, large model structure could not always be used in practical, especially for low cost devices where restricted hardware is used. This is on the contrary for Gaussian mixture model (GMM) where model parameters can be updated in real-time, but, with low modeling ability. In this paper, deeply integrated scheme combining these two models are proposed to improve adaptability and modeling ability. This is done by directly combining the results of models and feeding it back, together with the result of the DNN model, to update the GMM model. Besides, a control scheme is elaborately designed to detect the endpoints of speech. The superior performance by employing this scheme is validated through experiments in practical, which give an insight into the advantage of combining supervised learning and unsupervised learning.
\end{abstract}

\section{Introduction}
Voice detection from audio stream, namely voice activity detection (VAD), plays a key role in many fields, such as speech coding, speaker separation and recognition, speech recognition, etc \cite{VAD}. It is defined by extracting the starting and ending points of valid speech segment from a continuous audio signal. This will improve the data reliability after wiping off the invalid parts, thus, reducing the computational complexity and response time.

There are many algorithms to realize the detection, such as short-time energy, short-time average magnitude function (AM), short-time average zero-crossing rate (ZCR), short-time Auto Correlation, short-time average magnitude difference function (AMDF) in time domain, and Fourier analytics in frequency domain, etc \cite{algorithms}. All these conventional methods perform well in low noise environment. However, their performances are sensitive to environment and will degrade severely with interference. This makes them inappropriate in practice, especially for scenarios where complex noise exists, such as music, clicking, talking, coughing, etc \cite{performance}.

Recent years, deep neural network (DNN) is employed for VAD due to its superior modeling ability, especially for noisy environment \cite{DNN}. A robust model can be obtained if sufficient data especially for practical data with diversity can be used for training, together with a perfect model structure. However, this may not always be satisfied in practice, which will degrade the performance, and, this cannot be addressed by roughly increase the model complexity, especially for low cost devices where source-restricted hardware is used. In this context, a lite structure is more appropriate for implementation in practice, but in turn, the model adaptability will be reduced. On the contrary, the model parameters of Gaussian mixture model (GMM) can be updated in real-time with feedback structure, giving a higher adaptability to environment \cite{GMM}. Unfortunately, it has low modeling ability compared with that of DNN.

In this paper, a deeply integrated scheme combining these two models are proposed to improve adaptability and modeling ability. This is done by directly combining the results of models and feeding it back to update the GMM model. In the same time, the probabilities of speech and silence calculated by DNN model are used for parameters update of GMM model. Therefore, the attribution corresponding to speech or silence can be obtained for each frame. In addition, a control scheme is elaborately designed and illustrated to detect the endpoints of speech. The superior performance by employing this scheme is validated by experiments in practical.

\section{Algorithm Scheme}
\subsection{Algorithm structure}
\begin{figure}
\centering
\includegraphics[scale=0.55]{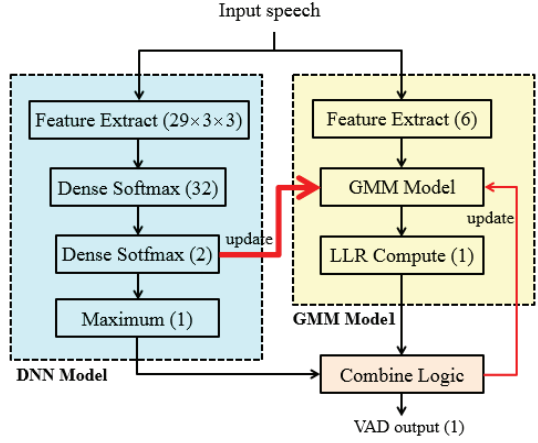}
\caption{Structure of combining DNN model with GMM model}
\label{fig:structure}
\end{figure}

The structure of the VAD algorithm is illustrated in Fig. \ref{fig:structure}. Two models namely DNN and GMM are used in the algorithm. For GMM model, GMM is used to model speech and noise respectively. For each input frame of audio, the probability of speech and that of noise are calculated respectively, and then the likelihood ratio of these two probabilities is calculated. The result is compared with a threshold value, and judged as speech if it is greater than the threshold value, otherwise is silence. The GMM can update the model parameters in real time according to environment. However, due to the limited modeling ability of GMM model, it is impossible to achieve accurate speech modeling, especially in complex audio environment, degrading the VAD performance. Fortunately, with superior modeling ability, the neural network (NN) can be employed to enhance the detection accuracy.

However, since the neural network is trained with corpus often constructed artificially, which is different from real data, there would be difference between the trained model and the real one. In addition, because data volume and data types cannot always be guaranteed to cover all real application scenarios in the training stage, there will also be some defects applying the NN model in practice. Therefore, the fusion of these two kinds of models can make use of the modeling ability of the neural network and the environmental adaptability of GMM model, so as to achieve better detection performance.

\subsection{DNN model}
\subsubsection{Feature extraction}
\begin{figure}
\centering
\includegraphics[scale=0.45]{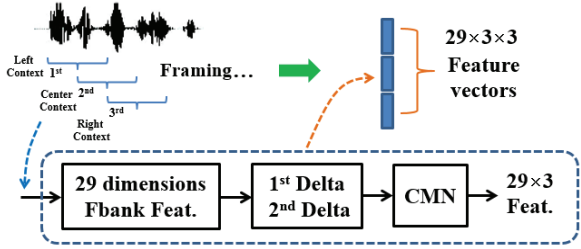}
\caption{Feature extraction for DNN model}
\label{fig:feature}
\end{figure}

For DNN model, 29 dimensions of Fbank features are first extracted and then differential between consecutive frames is performed to obtain the $1^{\rm{st}}$ and the $2^{\rm{nd}}$ delta features \cite{Delta}, followed by the cepstrum mean normalization (CMN) \cite{CMN}, resulting a dimension of $29 \times 3$ features. Thereafter, the features of the left context and the right context with respect to this frame are combined together for considering the temporal property of speech. These features are fed into a two layer dense neural network (DNN) with softmax activations as shown in Fig. \ref{fig:feature}. The number of nodes in hidden lay is 32 and that of the output is 2, where one is for the probability of speech and the other is for silence.

Since the accuracy of detection would be increasing with the complexity of DNN, only a lite two-layer DNN model is employed here for considering the hardware restriction and the response time of device, especially for lowcost devices.

\subsubsection{Data preparation}
Large volume of audio files with labels is required for DNN training. Two types of corpus can be manufactured and used for training. One is simulation data, and the other is practical data.

\begin{figure}
\centering
\includegraphics[scale=0.38]{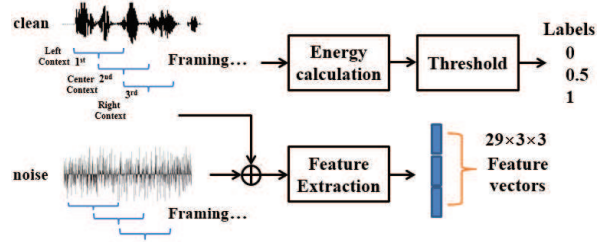}
\caption{Simulation data preparation}
\label{fig:simulation}
\end{figure}

$\bf{Simulation\ data}$. This is obtained by adding different type of noise to the clean files and calculate the feature vectors, while the label corresponding to this noisy frame is generated by calculating the energy of the clean speech frame and comparing with the threshold as depicted in Fig. \ref{fig:simulation}. Aside from the actual labels of ``0'' and ``1'', an additional label ``0.5'' is employed for considering the uncertainty of speech during transitions.

\begin{figure}
\centering
\includegraphics[scale=0.50]{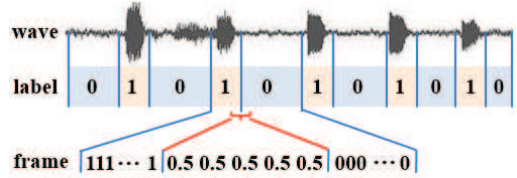}
\caption{Label design of practical data}
\label{fig:practice}
\end{figure}

$\bf{Practical\ data}$. It is obtained by collecting them from the real devices. For generalization, different types of device with different performance of hardware and software algorithm should be included for diversity. Therefore, the work lays on the manipulation of labels corresponding to this corpus. Fortunately, this can be done by employing annotation tool such as Praat and etc. As depicted in Fig. \ref{fig:practice}, the long segments belong to speech or noise can be easily obtained with annotation tool. Since these labels are obtained roughly, a refined label corresponding to each should be prepared. This can be done by partitioning the long segment into frames with same labels. As is noted above, the labels around the boundary could be replaced by ``0.5'' for considering the uncertainty of these regions.

\subsection{GMM model}
The GMM model from open source of Web Real-Time Communication (WebRTC) \cite{WebRTC, WebRTC1, WebRTC2} is employed and schematically derived here. With the independent assumption for this two mixture components, i.e., ${p_s}\left( {\mathord{\buildrel{\lower3pt\hbox{$\scriptscriptstyle\rightharpoonup$}}
\over x} ,\mathord{\buildrel{\lower3pt\hbox{$\scriptscriptstyle\rightharpoonup$}}
\over y} ;i} \right) = {p_s}\left( {\mathord{\buildrel{\lower3pt\hbox{$\scriptscriptstyle\rightharpoonup$}}
\over x} ;i} \right) \cdot {p_s}\left( {\mathord{\buildrel{\lower3pt\hbox{$\scriptscriptstyle\rightharpoonup$}}
\over y} ;i} \right)$ the log-likelihood ratio (LLR) of the GMM model of the WebRTC can be expressed by
\begin{equation}
\begin{array}{l}
{L_i}\left( {\mathord{\buildrel{\lower3pt\hbox{$\scriptscriptstyle\rightharpoonup$}}
\over \varphi } ;i} \right) \buildrel \Delta \over = {L_i}\left( {\mathord{\buildrel{\lower3pt\hbox{$\scriptscriptstyle\rightharpoonup$}}
\over x} ,\mathord{\buildrel{\lower3pt\hbox{$\scriptscriptstyle\rightharpoonup$}}
\over y} ;i} \right) = \log \left( {\frac{{{p_s}\left( {\mathord{\buildrel{\lower3pt\hbox{$\scriptscriptstyle\rightharpoonup$}}
\over x} ,\mathord{\buildrel{\lower3pt\hbox{$\scriptscriptstyle\rightharpoonup$}}
\over y} ;i} \right)}}{{{p_n}\left( {\mathord{\buildrel{\lower3pt\hbox{$\scriptscriptstyle\rightharpoonup$}}
\over x} ,\mathord{\buildrel{\lower3pt\hbox{$\scriptscriptstyle\rightharpoonup$}}
\over y} ;i} \right)}}} \right)\\
\begin{array}{*{20}{c}}
{}&{}&{}
\end{array} \approx \frac{{\exp \left( { - \frac{{{{\left( {\mathord{\buildrel{\lower3pt\hbox{$\scriptscriptstyle\rightharpoonup$}}
\over x}  - {u_{\mathord{\buildrel{\lower3pt\hbox{$\scriptscriptstyle\rightharpoonup$}}
\over x} ,s}}} \right)}^2}}}{{\sigma _{\mathord{\buildrel{\lower3pt\hbox{$\scriptscriptstyle\rightharpoonup$}}
\over x} ,s}^2}}} \right) + \exp \left( { - \frac{{{{\left( {\mathord{\buildrel{\lower3pt\hbox{$\scriptscriptstyle\rightharpoonup$}}
\over y}  - {u_{\mathord{\buildrel{\lower3pt\hbox{$\scriptscriptstyle\rightharpoonup$}}
\over y} ,s}}} \right)}^2}}}{{\sigma _{\mathord{\buildrel{\lower3pt\hbox{$\scriptscriptstyle\rightharpoonup$}}
\over y} ,s}^2}}} \right)}}{{\exp \left( { - \frac{{{{\left( {\mathord{\buildrel{\lower3pt\hbox{$\scriptscriptstyle\rightharpoonup$}}
\over x}  - {u_{\mathord{\buildrel{\lower3pt\hbox{$\scriptscriptstyle\rightharpoonup$}}
\over x} ,n}}} \right)}^2}}}{{\sigma _{\mathord{\buildrel{\lower3pt\hbox{$\scriptscriptstyle\rightharpoonup$}}
\over x} ,n}^2}}} \right) + e\left( { - \frac{{{{\left( {\mathord{\buildrel{\lower3pt\hbox{$\scriptscriptstyle\rightharpoonup$}}
\over y}  - {u_{\mathord{\buildrel{\lower3pt\hbox{$\scriptscriptstyle\rightharpoonup$}}
\over y} ,n}}} \right)}^2}}}{{\sigma _{\mathord{\buildrel{\lower3pt\hbox{$\scriptscriptstyle\rightharpoonup$}}
\over y} ,n}^2}}} \right)}}
\end{array}
  \label{eq1}
\end{equation}
where $\rm{exp}(x)$ denotes the exponential function of $x$,   $\mathord{\buildrel{\lower3pt\hbox{$\scriptscriptstyle\rightharpoonup$}}
\over s}  = \left( {\mathord{\buildrel{\lower3pt\hbox{$\scriptscriptstyle\rightharpoonup$}}
\over x} ,\mathord{\buildrel{\lower3pt\hbox{$\scriptscriptstyle\rightharpoonup$}}
\over y} } \right)$ is the input feature vectors with two mixture components $\mathord{\buildrel{\lower3pt\hbox{$\scriptscriptstyle\rightharpoonup$}}
\over x}$ and $\mathord{\buildrel{\lower3pt\hbox{$\scriptscriptstyle\rightharpoonup$}}
\over y}$, ${\mathord{\buildrel{\lower3pt\hbox{$\scriptscriptstyle\rightharpoonup$}}
\over u} _{\mathord{\buildrel{\lower3pt\hbox{$\scriptscriptstyle\rightharpoonup$}}
\over x} ,s}}$ and $\sigma _{\mathord{\buildrel{\lower3pt\hbox{$\scriptscriptstyle\rightharpoonup$}}
\over x} ,s}^2$ are the mean and variance values of a subband of speech corresponding to variable $\mathord{\buildrel{\lower3pt\hbox{$\scriptscriptstyle\rightharpoonup$}}
\over x}$, ${\mathord{\buildrel{\lower3pt\hbox{$\scriptscriptstyle\rightharpoonup$}}
\over u} _{\mathord{\buildrel{\lower3pt\hbox{$\scriptscriptstyle\rightharpoonup$}}
\over y} ,s}}$ and $\sigma _{\mathord{\buildrel{\lower3pt\hbox{$\scriptscriptstyle\rightharpoonup$}}
\over y} ,s}^2$ mean and variance values of a subband of speech corresponding to variable $\mathord{\buildrel{\lower3pt\hbox{$\scriptscriptstyle\rightharpoonup$}}
\over y}$. ${\mathord{\buildrel{\lower3pt\hbox{$\scriptscriptstyle\rightharpoonup$}}
\over u} _{\mathord{\buildrel{\lower3pt\hbox{$\scriptscriptstyle\rightharpoonup$}}
\over x} ,n}}$ and ${\mathord{\buildrel{\lower3pt\hbox{$\scriptscriptstyle\rightharpoonup$}}
\over u} _{\mathord{\buildrel{\lower3pt\hbox{$\scriptscriptstyle\rightharpoonup$}}
\over y} ,n}}$ are the mean values of a subband of noise and $\begin{array}{*{20}{c}}
{\sigma _{\mathord{\buildrel{\lower3pt\hbox{$\scriptscriptstyle\rightharpoonup$}}
\over x} ,n}^2}
\end{array}$ and $\begin{array}{*{20}{c}}
{\sigma _{\mathord{\buildrel{\lower3pt\hbox{$\scriptscriptstyle\rightharpoonup$}}
\over y} ,n}^2}
\end{array}$ are the corresponding variances. The Gaussian distribution is denoted by
$
p\left( {\mathord{\buildrel{\lower3pt\hbox{$\scriptscriptstyle\rightharpoonup$}}
\over x} } \right) = \frac{1}{{\sqrt {2\pi } \sigma }}{e^{ - \frac{{{{\left( {\mathord{\buildrel{\lower3pt\hbox{$\scriptscriptstyle\rightharpoonup$}}
\over x}  - \mathord{\buildrel{\lower3pt\hbox{$\scriptscriptstyle\rightharpoonup$}}
\over u} } \right)}^2}}}{{2{\sigma ^2}}}}}
$.

There are six subands considered in WebRTC. The overall LLR is obtained by weighting average that of these subbands as
\begin{equation}
{L_t}\left( {\mathord{\buildrel{\lower3pt\hbox{$\scriptscriptstyle\rightharpoonup$}}
\over x} ,\mathord{\buildrel{\lower3pt\hbox{$\scriptscriptstyle\rightharpoonup$}}
\over y} } \right) = \sum\limits_{i = 1}^6 {{k_i}{L_i}\left( {\mathord{\buildrel{\lower3pt\hbox{$\scriptscriptstyle\rightharpoonup$}}
\over x} ,\mathord{\buildrel{\lower3pt\hbox{$\scriptscriptstyle\rightharpoonup$}}
\over y} ;i} \right)}
  \label{eq2}
\end{equation}

The LLR of each subband is first compared with threshold ${T_\tau}$ to decide whether it is speech or not. Speech is considered if one of the subbands is confirmed. Otherwise, the overall LLR is compared with the threshold ${T_a}$ by
\begin{equation}
{F_{{\rm{vad}}}} = \left\{ {\begin{array}{*{20}{c}}
{1,}\\
{0,}
\end{array}} \right.\begin{array}{*{20}{c}}
{\left. {{L_i} > {T_\tau }} \right\|{L_t} > {T_a}}\\
{{\rm{otherwise}}}
\end{array}
  \label{eq3}
\end{equation}

The update equations of mean and variance for noise is summarized as
\begin{equation}
\begin{array}{l}
{\nabla _{{u_{nj}}}} = \frac{{x - {u_{nj}}}}{{\sigma _{nj}^2}},\begin{array}{*{20}{c}}
{}
\end{array}{\nabla _{{\sigma _{nj}}}} = \frac{1}{{{\sigma _{nj}}}}\left( {\frac{{{{\left( {x - {u_{nj}}} \right)}^2}}}{{\sigma _{nj}^2}} - 1} \right)\\
{u_{nj}}\left( n \right) = {u_{nj}}\left( {n - 1} \right) + \left( {1 - {F_{{\rm{vad}}}}\left( n \right)} \right) \cdot {K_{\Delta n}} \cdot {\nabla _{{u_{nj}}}}\\
\begin{array}{*{20}{c}}
{}&{}&{}
\end{array}\begin{array}{*{20}{c}}
{}
\end{array} \cdot \frac{{{p_n}\left( {\left. {x\left( n \right)} \right|{H_0}} \right)}}{{{p_n}\left( {\left. {x\left( n \right)} \right|{H_0}} \right) + {p_n}\left( {\left. {x\left( n \right)} \right|{H_1}} \right)}} + {K_L}\left[ {{x_{\min }}\left( n \right) - {u_{nj}}\left( n \right)} \right]\\
{\sigma _{nj}}\left( n \right) = {\sigma _{nj}}\left( {n - 1} \right) + \left( {1 - {F_{{\rm{vad}}}}\left( n \right)} \right) \cdot {C_{\Delta n}} \cdot {\nabla _{{\sigma _{nj}}}}\\
\begin{array}{*{20}{c}}
{}&{}&{}
\end{array}\begin{array}{*{20}{c}}
{}
\end{array} \cdot \frac{{{p_n}\left( {\left. {x\left( n \right)} \right|{H_0}} \right)}}{{{p_n}\left( {\left. {x\left( n \right)} \right|{H_0}} \right) + {p_n}\left( {\left. {x\left( n \right)} \right|{H_1}} \right)}}
\end{array}
  \label{eq4}
\end{equation}

Similarly, the update equations of mean and variance for speech is summarized as
\begin{equation}
\begin{array}{l}
{\nabla _{{u_{sj}}}} = \frac{{x - {u_{sj}}}}{{\sigma _{sj}^2}},\begin{array}{*{20}{c}}
{}
\end{array}{\nabla _{{\sigma _{sj}}}} = \frac{1}{{{\sigma _{sj}}}}\left( {\frac{{{{\left( {x - {u_{sj}}} \right)}^2}}}{{\sigma _{sj}^2}} - 1} \right)\\
{u_{sj}}\left( n \right) = {u_{sj}}\left( {n - 1} \right) + {F_{{\rm{vad}}}}\left( n \right) \cdot {K_{\Delta n}} \cdot {\nabla _{{u_{nj}}}}\\
\begin{array}{*{20}{c}}
{}&{}&{}
\end{array}\begin{array}{*{20}{c}}
{}
\end{array} \cdot \frac{{{p_n}\left( {\left. {x\left( n \right)} \right|{H_1}} \right)}}{{{p_n}\left( {\left. {x\left( n \right)} \right|{H_0}} \right) + {p_n}\left( {\left. {x\left( n \right)} \right|{H_1}} \right)}}\\
{\sigma _{sj}}\left( n \right) = {\sigma _{1j}}\left( {n - 1} \right) + {F_{{\rm{vad}}}}\left( n \right) \cdot {C_{\Delta n}} \cdot {\nabla _{{\sigma _{sj}}}}\\
\begin{array}{*{20}{c}}
{}&{}&{}
\end{array}\begin{array}{*{20}{c}}
{}
\end{array} \cdot \frac{{{p_n}\left( {\left. {x\left( n \right)} \right|{H_1}} \right)}}{{{p_n}\left( {\left. {x\left( n \right)} \right|{H_0}} \right) + {p_n}\left( {\left. {x\left( n \right)} \right|{H_1}} \right)}}
\end{array}
  \label{eq6}
\end{equation}
where the subscript $n$ denotes noise, $j$ denotes the $j$-th component of GMM, and the coefficients are chosen as ${K_{\Delta n}} = 0.02$, ${K_{\Delta s}} = 0.2$, ${C_{\Delta n}} = 0.1$, ${K_L} = 0.6$ in WebRTC. The minimum value of noise denoted by ${x_{\min }}\left( n \right)$ is obtained by
\begin{equation}
{x_{\min }}\left( n \right) = \left\{ {\begin{array}{*{20}{c}}
{\left\{ \begin{array}{l}
\begin{array}{*{20}{c}}
{\rm{if},}
\end{array}{x_{\min }}\left( n \right) > {x_{\min }}\left( {n - 1} \right)\\
\rm{then},\begin{array}{*{20}{c}}
{0.99 \cdot {x_{\min }}\left( {n - 1} \right) + 0.01 \cdot {u_n}\left( {n - 1} \right)}
\end{array}
\end{array} \right.}\\
{\left\{ \begin{array}{l}
\begin{array}{*{20}{c}}
{\rm{if},}
\end{array}{x_{\min }}\left( n \right) < {x_{\min }}\left( {n - 1} \right)\\
\rm{then},\begin{array}{*{20}{c}}
{0.20 \cdot {x_{\min }}\left( {n - 1} \right) + 0.80 \cdot {u_n}\left( {n - 1} \right)}
\end{array}
\end{array} \right.}
\end{array}} \right.
  \label{eq8}
\end{equation}
In this way, the noise would increase slowly for trying to keep the primary noise invariant, while decrease rapidly to keep tracking the variation of noise.

\subsection{Combination scheme}
The two outputs corresponding to the probabilities of speech and silence from DNN can be used for optimizing the model updating of GMM, and the judgments of these two models can be combined together by
\begin{equation}
{F_{{\rm{vad}}}} = \left\{ {\begin{array}{*{20}{c}}
{1,}\\
{F_{{\rm{vad}}}^{{\rm{GMM}}},}
\end{array}\begin{array}{*{20}{c}}
{F_{{\rm{vad}}}^{{\rm{DNN}}} = 1}\\
{F_{{\rm{vad}}}^{{\rm{DNN}}} = 0}
\end{array}} \right.
  \label{eq9}
\end{equation}
where $F_{{\rm{vad}}}^{{\rm{DNN}}}$ and $F_{{\rm{vad}}}^{{\rm{GMM}}}$ are the detection results of the DNN and GMM models respectively. This result is then be used for controlling the parameters updating of GMM model as indicated by Eq. (\ref{eq4}) $\sim$ Eq. (\ref{eq6}).

Moreover, the probability of the noise and that of speech can also the updated and optimized by utilizing the softmax output of DNN model. This can be done by the equations bellow.

\begin{equation}
\begin{array}{l}
{p_n}\left( {\left. {x\left( n \right)} \right|{H_0}} \right) = \underbrace {\alpha  \cdot {p_n}\left( {n - 1} \right)}_{{\rm{DNN\ model}}} + \underbrace {\left( {1 - \alpha } \right) \cdot {p_n}\left( {\left. {x\left( {n - 1} \right)} \right|{H_0}} \right)}_{{\rm{GMM\ model}}}\\
{p_n}\left( {\left. {x\left( n \right)} \right|{H_1}} \right) = \underbrace {\beta  \cdot {p_s}\left( {n - 1} \right)}_{{\rm{DNN\ model}}} + \underbrace {\left( {1 - \beta } \right) \cdot {p_n}\left( {\left. {x\left( {n - 1} \right)} \right|{H_1}} \right)}_{{\rm{GMM\ model}}}\\
\end{array}
  \label{eq10}
\end{equation}
where ${p_n}\left( {n - 1} \right)$ and ${p_s}\left( {n - 1} \right)$ are the probabilities corresponding to noise and speech of the ($n-1$)-th frame of the DNN model, ${p_n}\left( {\left. {x\left( {n - 1} \right)} \right|{H_0}} \right)$ and ${p_n}\left( {\left. {x\left( {n - 1} \right)} \right|{H_1}} \right)$ are those of the GMM model, $\alpha$ and $\beta$ are the smoothing coefficients. Moreover, the probabilities should be normalized.

The values of smoothing coefficients can be selected based on the fact that the DNN has superior ability to modeling speech than that of GMM, while, it is more likely to be noise if determined by GMM model due to the rigorous judgement of noise. Thus, the coefficients can be selected as $\alpha  = 0.1, \  \beta  = 0.8$.

\section{Endpoints Detection}
\begin{figure}[H]
\centering
\includegraphics[scale=0.45]{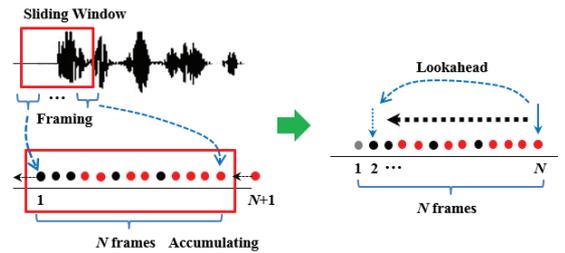}
\caption{Schematic design of endpoint detection}
\label{fig:detection}
\end{figure}

With the aforementioned calculation model, the attribution of each frame can be determined to be speech or not. In this context, a judging logic to calculate the beginning and the ending, i.e., the endpoints of speech, is required as depicted in Fig. \ref{fig:detection}.

A sliding window is used for accumulate the number of frames corresponding to speech as shown in Fig. \ref{fig:detection} where each point is corresponding to a frame and is marked by a specific color representing whether it is speech or not. Here, the one marked by red is considered as speech and the other is silence. The window is sliding with time, that is, every time one frame is sliding out and another one is coming in. Only when the number of speech frames in the $N$ frames window is larger than the threshold denoted by $\rho  \cdot N$ (where $\rho$ is the percentage of speech frames with respect to the window length $N$), would endpoint of speech be detected. However, this is not the real endpoint considering the durative property of speech. In this context, the real one can be obtained by looking back with $M$ frames as shown in Fig. \ref{fig:detection} where the position is traceback from the $N$-th frame to the $2^{\rm{nd}}$ frame. It is notable that, the parameter $M$ could be larger than the window length $N$ only when the frame marked 1 is not the real $1^{\rm{st}}$ frame of the long speech.

As long as the endpoints are obtained, would the speech segment be extracted out. In practice, a data buffer with finite length would be used for reserving the incoming data during detection. There may be some situations that should be taken into consideration in terms of the length of buffer and the segment length of detected speech.

\subsection{Inside data buffer}
\begin{figure}
\centering
\includegraphics[scale=0.8]{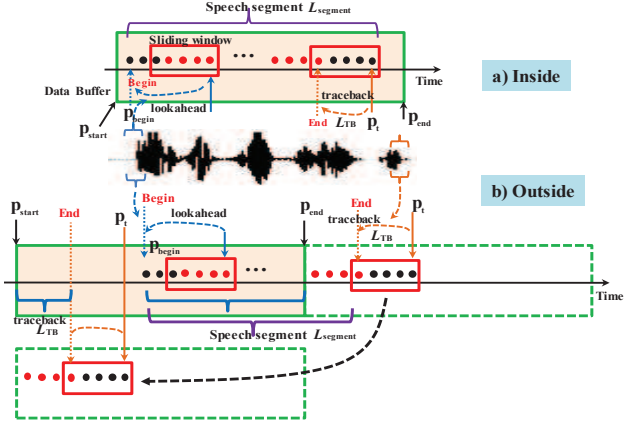}
\caption{Speech segment overflows data buffer}
\label{fig:across}
\end{figure}

When the length of the detected speech segment is within the length of data buffer, the pointers corresponding to the beginning and ending of the speech segment can be directly obtained as depicted in Fig. \ref{fig:across}(a) by
\begin{equation}
{{\rm{P}}_{{\rm{out}}}} = {{\rm{P}}_{{\rm{begin}}}},{L_{{\rm{out}}}} = {L_{{\rm{segment}}}} - {L_{{\rm{TB}}}}
\label{eq12}
\end{equation}
where ${{\rm{P}}_{\rm{t}}}$ and ${{\rm{P}}_{{\rm{begin}}}}$ are the pointers of the current frame and the detected beginning of the speech segment, ${L_{{\rm{segment}}}}$ and ${L_{{\rm{TB}}}}$ are the length of the available speech segment and the traceback. ${L_{{\rm{segment}}}}$ is obtained by accumulating the frames as long as the beginning of the speech segment is detected. Therefore, with the beginning pointer ${{\rm{P}}_{{\rm{out}}}}$ and the data length ${L_{{\rm{out}}}}$, the speech segment can be extracted.


\subsection{Outside data buffer}
When the end point of the speech segment exceeds the current data buffer as depicted in Fig. \ref{fig:across}(b), these overflowed data will be rolled back to the front of the data buffer and cover the old data. Three additional cases would occur as follows.

a) The end point of speech segment is pulled back to the position within the data buffer due to traceback. Therefore, the endpoints of the speech segment can be directly written as
\begin{equation}
\begin{array}{l}
\begin{array}{*{20}{c}}
{{\rm{if}}\begin{array}{*{20}{c}}
{\left( {{{\rm{P}}_{\rm{t}}} - {{\rm{P}}_{{\rm{start}}}} < {L_{{\rm{TB}}}}} \right)}
\end{array}}
\end{array}\\
{\rm{then}}\begin{array}{*{20}{c}}
{{{\rm{P}}_{{\rm{out,1}}}} = {{\rm{P}}_{{\rm{begin}}}},{L_{{\rm{out,1}}}} = {L_{{\rm{segment}}}} - {L_{{\rm{TB}}}}}
\end{array}
\end{array}
\label{eq13}
\end{equation}

b) The end point of speech segment is indeed be rolled back to the front of the data buffer but is at the left side of the beginning point of the speech segment. In this context, the speech segment is partitioned into two parts, and the beginning and ending of the speech segment can be expressed as
\begin{equation}
\begin{array}{l}
\begin{array}{*{20}{c}}
{{\rm{if}}\begin{array}{*{20}{c}}
{\left( {{{\rm{P}}_{\rm{t}}} - {{\rm{P}}_{{\rm{start}}}} > {L_{{\rm{TB}}}}} \right)}
\end{array}}
\end{array}\\
{\rm{then}}\left\{ \begin{array}{l}
{{\rm{P}}_{{\rm{out,1}}}} = {{\rm{P}}_{{\rm{begin}}}},{L_{{\rm{out,1}}}} = {{\rm{P}}_{{\rm{end}}}} - {{\rm{P}}_{{\rm{begin}}}}\\
{{\rm{P}}_{{\rm{out,2}}}} = {{\rm{P}}_{{\rm{start}}}},{L_{{\rm{out,2}}}} = \left( {{{\rm{P}}_{\rm{t}}} - {{\rm{P}}_{{\rm{start}}}}} \right) - {L_{{\rm{TB}}}}
\end{array} \right.
\end{array}
\label{eq14}
\end{equation}
The overall speech segment is obtained by concatenating the pointer of ${{\rm{P}}_{{\rm{out,1}}}}$ and ${{\rm{P}}_{{\rm{out,2}}}}$.

c) The end point has not been detected even when the pointer is overflowed and rolled back to the front of the data buffer. Moreover, the pointer is now just at the left side of the beginning of the speech segment, if the data reserved in the buffer is not taken away, it will be covered by the new coming data. Therefore, all the data from the beginning of the speech should be copy away immediately no matter whether the end point has been detected or not. The pointer can be expressed as
\begin{equation}
\begin{array}{l}
\begin{array}{*{20}{c}}
{{\rm{if}}\begin{array}{*{20}{c}}
{\left( {{L_{{\rm{segment}}}} \ge {L_{{\rm{buffer}}}}} \right)}
\end{array}}
\end{array}\\
{\rm{then}}\begin{array}{*{20}{c}}
{\left\{ \begin{array}{l}
{{\rm{P}}_{{\rm{out,1}}}} = {{\rm{P}}_{{\rm{begin}}}},{L_{{\rm{out,1}}}} = {{\rm{P}}_{{\rm{end}}}} - {{\rm{P}}_{{\rm{begin}}}}\\
{{\rm{P}}_{{\rm{out,2}}}} = {{\rm{P}}_{{\rm{start}}}},{L_{{\rm{out,2}}}} = {{\rm{P}}_{{\rm{begin}}}} - {{\rm{P}}_{{\rm{start}}}}
\end{array} \right.}
\end{array}
\end{array}
\label{eq15}
\end{equation}


\section{Validations}
The performance is validated from the perspective of experiments and simulations.

$\bf{Experiments}$. The proposed scheme is validated on four platforms for comparison, each equipped with different algorithms. Due to hardware restriction, the enhancement algorithms on the devices are different, thus, giving different quality of speech. Moreover, the parameters for detecting the beginning and the ending of speech segment are tuned to obtain an optimal accuracy with the same length of data buffer. Due to hardware restrictions, the VAD model used for devices named ``A'' and ``B'' are GMM model with two mixture components. While, it is DNN model for devices named ``C'' and ``D'' which are high-performance hardware.  As is depicted in Table~\ref{tab:accuracy}, it has superior detecting accuracy than the standalone scheme.

$\bf{Simulations}$. The detection accuracy at different noisy conditions are compared among the algorithms shown in Table \ref{tab:accuracy1}. This is done by adding different type of noise into the clean speech with different signal-to-noise ratio (SNR). It can be seen that due to insufficient training, VAD with DNN model could be inferior than that with GMM model. While, it is obvious that the proposed scheme is superior to the standalone models.

\begin{table}[H]
\centering
\caption{Comparisons by experiments}
  \label{tab:accuracy}
 \begin{tabular}{|c|c|c|c|c|c|}
   \hline
   \multirow{2}{*}{\bf{Algorithm}}  &   \multicolumn{4}{c|}{\bf{Device Type}} \\
   \cline{2-5}
    & $\rm{A}$  & $\rm{B}$  & $\rm{C}$  & $\rm{D}$ \\
   \hline
   GMM  & 90.74$\%$  & $86.67\%$  & $\diagdown$  & $\diagdown$ \\
   \hline
   DNN  & $\diagdown$ & $\diagdown$ & $77.55\%$  & 98$\%$ \\
   \hline
   Proposed  & 92.59$\%$ & $94.29\%$ & $97.01\%$ & 100$\%$ \\
   \hline
\end{tabular}
\end{table}

\begin{table}[H]
\centering
\caption{Comparisons by simulations}
  \label{tab:accuracy1}
 \begin{tabular}{|c|c|c|c|c|c|}
   \hline
    \multirow{2}{*}{\bf{SNR}} & \multirow{2}{*}{\bf{Noise Type}} &  \multicolumn{3}{c|}{\bf{Algorithm}} \\ \cline{3-5}

    & & $\rm{GMM}$  & $\rm{DNN}$  & $\rm{Proposed}$  \\ \hline

    \multirow{4}{*}{\bf{5dB}}
    & Wind        & 95.2$\%$   & 97.60$\%$  & 100$\%$  \\ \cline{2-5}
    & Water       & 97.6$\%$   & 40.40$\%$  & 100$\%$  \\ \cline{2-5}
    & Babble      & 90.40$\%$  & 81.00$\%$  & 100$\%$  \\ \cline{2-5}
    & Television  & 95.20$\%$  & 92.80$\%$  & 97.60$\%$  \\ \hline

    \multirow{4}{*}{\bf{10dB}}
    & Wind       & 97.60$\%$  & 100$\%$    & 100$\%$  \\ \cline{2-5}
    & Water      & 97.60$\%$  & 69$\%$     & 100$\%$  \\ \cline{2-5}
    & Babble     & 95.20$\%$  & 97.60$\%$  & 100$\%$  \\ \cline{2-5}
    & Television & 100$\%$    & 97.60$\%$  & 100$\%$  \\ \hline

    \multirow{4}{*}{\bf{15dB}}
    & Wind       & 97.60$\%$  & 100$\%$    & 100$\%$  \\ \cline{2-5}
    & Water      & 100$\%$    & 97.60$\%$  & 100$\%$  \\ \cline{2-5}
    & Babble     & 97.60$\%$  & 97.60$\%$  & 100$\%$  \\ \cline{2-5}
    & Television & 100$\%$    & 100$\%$    & 100$\%$  \\ \hline

\end{tabular}
\end{table}

\section{Conclusions}
A deeply integrated scheme combining DNN and GMM models is proposed in this paper to estimate whether the frame is speech or not. The modeling ability of DNN model and the adaptive ability of GMM model can be utilized together for improving the estimation accuracy. With the lite structure of the integrated scheme, it is more suitable for implementation in practice. In addition to the estimation algorithm, a detecting scheme to extract the speech segment from an audio stream is elaborately designed and detailed discussed. The proposed scheme is validated from practical experiments, demonstrating superior performance. This provides a insight into the advantage for combining supervised learning and unsupervised learning.

\bibliographystyle{IEEEtran}

\bibliography{mybib}

\begin{thebibliography}{1}
\providecommand{\url}[1]{#1}
\csname url@samestyle\endcsname
\providecommand{\newblock}{\relax}
\providecommand{\bibinfo}[2]{#2}
\providecommand{\BIBentrySTDinterwordspacing}{\spaceskip=0pt\relax}
\providecommand{\BIBentryALTinterwordstretchfactor}{4}
\providecommand{\BIBentryALTinterwordspacing}{\spaceskip=\fontdimen2\font plus
\BIBentryALTinterwordstretchfactor\fontdimen3\font minus
  \fontdimen4\font\relax}
\providecommand{\BIBforeignlanguage}[2]{{%
\expandafter\ifx\csname l@#1\endcsname\relax
\typeout{** WARNING: IEEEtran.bst: No hyphenation pattern has been}%
\typeout{** loaded for the language `#1'. Using the pattern for}%
\typeout{** the default language instead.}%
\else
\language=\csname l@#1\endcsname
\fi
#2}}
\providecommand{\BIBdecl}{\relax}
\BIBdecl

\bibitem{VAD}
D.~Singh, F.~Boland, ``Voice activity detection,''
\emph{Crossroads}, vol.~13, no.~4, pp.~7, 2007.

\bibitem{algorithms}
X.~Yang, B.~Tan, J.~Ding, and et~al., ``Comparative study on voice activity detection algorithm,''
in \emph{International Conference on Electrical and Control Engineering}, iCECE, 2010, pp.~599--602.

\bibitem{performance}
C.~K.~Pham, ``Noise robust voice activity detection,'' Master dissertation,
Nanyang Technological University, 2012

\bibitem{DNN}
X.~L.~Zhang, J.~WU, ``Deep belief network based voice activity detection,''
\emph{IEEE Transactions on Audio, Speech, and Language Processing}, vol.~21, no.~4, pp.~691--710, 2013.

\bibitem{GMM}
D.~Ying, Y.~Yan, J.~Dang, et~al., ``Voice activity detection based on an unsupervised learning framework,''
\emph{IEEE Transactions on Audio Speech and Language Processing}, vol.~19, no.~8, pp.~2624--2633, 2011.

\bibitem{Delta}
M.~Chen, X.~He, J.~Yang and et~al., ``3-D convolutional recurrent neural networks with attention model for speech emotion recognition,''
\emph{IEEE Signal Processing Letters}, vol.~25, no.~10, pp.~1440--1444, 2018.

\bibitem{CMN}
M.~Morishima, T.~Isobe, J.~Takahashi, ``Phonetically adaptive cepstrum mean normalization for acoustic mismatch compensation,''
\emph{IEEE Workshop on Automatic Speech Recognition $\&$ Understanding}, ASRU, 1997, pp.~436-441.

\bibitem{WebRTC}
R.~Chand, H.~prasad, ``Real time communication using WebRTC,''
\emph{International Journal of Mobile Computing and Application}, vol.~5. vol.~1--3, 2018.

\bibitem{WebRTC1}
Salvatore Loreto, Simon Pietro Romano, ``Real-time communication with WebRTC,'' O'Reilly Media, Inc., 2014.

\bibitem{WebRTC2}
WebRTC Source, https://chromium.googlesource.com/external/webrtc.

\end{thebibliography}


\end{document}